\documentstyle[11pt,aaspp4,epsf]{article}

\begin{document}

\title{On the Theoretical Period-Luminosity Relation of Cepheids} 
       
\author{ Hideyuki Saio}
\affil{Astronomical Institute, School of Science,
Tohoku University, 980 Sendai, Japan; saio@astr.tohoku.ac.jp}

\and 

\author{Alfred Gautschy}
\affil{Astronomisches Institut der Universit\"at Basel,
       Venusstr.~7, 4102 Binningen, Switzerland; gautschy@astro.unibas.ch}


\begin{abstract}
	The theoretical period-luminosity relation for Cepheids was
        deduced from stellar evolution and pulsation computations.
        Questions concerning the metallicity dependence of the
        relation as well as the constancy of the pulsation `constant'
        $Q$ in the instablility strip 
        were addressed. We found no significant metallicity
        dependence of either the slope or of the zero-point of the
        period-luminosity relation. The $Q$-value, however, varies
        across the instability strip, in effective temperature as well
        as in luminosity; its metallicity dependence, on the other
        hand, is marginal.
\end{abstract}

\keywords{stars: Cepheids --- stars: evolution --- stars: pulsation}

\section{Introduction}

The most influential role Cepheid variables play in galactic and
extragalactic astronomy is through their period-luminosity (PL)
relation.  The rather easy measurement of the period of their regular
pulsation allows to estimate an absolute magnitude or luminosity and
therefore to derive eventually the distance to the variable star. The
high luminosities of Cepheids favors them to anchor the extragalactic
distance scale. Due to the still debated value of the Hubble constant,
a secure calibration and a solid understanding of the PL relation is
of fundamental importance. References to the enormous body of
literature on the PL relation and directions to historic accounts can
be found e.g. in Cox (1980), Madore and Freedman (1991), Gieren and
Fouqu{\'e} (1993).

Theoretical PL relations of Cepheids were commonly determined by
referring to mass-luminosity (ML) relations which were derived from
stellar evolution computations. Thereupon separate stellar envelope
models were calculated whose stability properties were eventually
studied (e.g. \cite{bit77}, \cite{ir84}, \cite{sto88},
\cite{cwc93}). The authors are not aware of any PL-relation studies in
which stability properties of stellar evolution models were computed
directly.

In terms of recent theoretical work on the PL relation of Cepheids we
should mention Stothers (1988). He used various published sources of
stellar evolution computations with differing composition and
different physical assumptions to derive ML and PL relations. The
periods of the radial fundamental modes were already parameterized
through a period-mass-radius relation which was fitted through
numerical data. The aim of Stother's study was to quantify the
relative dependences of the PL and period-luminosity-color (PLC)
relation on chemical composition on purely theoretical grounds.
Later, Chiosi et al. (1993) performed pulsation computations on
envelope models which were calculated on the basis of ML relations
derived from stellar evolution tracks from the literature and from a
few unpublished computations. A range of chemical compositions which
bracketed Galactic as well as Magellanic Cepheids were considered. The
effect of overshooting was included in the modeling; for mild
overshooting $\log L$ of the of the ML relation increased by 0.25 and
it was shifted by $\log L$ by 0.5 for full overshooting.  Eventually,
analytical fits to the numerical data were transformed into
astronomical observables (colors and magnitudes in different
band-passes). The authors favored models with considerable overshoot
over ones neglecting it as the former ones apparently solved the
long-standing mass discrepancy problem for the Cepheids.  This
particular problem was, however, considerably defused by the
application of the new generation of opacity data
(\cite{mbm92}). Therefore, the question concerning a realistic
overshooting length of convective elements at the edge of the convective
core has to be re-addressed from a different viewpoint.

Due to recent high-precision observations with HST and Hipparchos
accuracy considerations of the distance-scale calibration relying on
Cepheids gained new momentum (\cite{bms96}, \cite{fc97}, \cite{st97}).
However, suggestions of the PL relation being sensitive to metallicity
effects (for an extreme viewpoint see \cite{sefuk97}) overshadowed the
distance determination of galaxies with different mean metallicities.
A solid understanding of the theoretical basis of the PL relation is
indispensable (Sandage 1996) to prevent possible systematic 
misinterpretation of the data.  

This paper presents self-consistently derived blue-edges of radial
fundamental modes of model stars which were obtained from evolution
computations. Heavy-element and helium abundances were varied and
evolutionary tracks of stars with various masses were recomputed to
find possible shifts of the location of the blue edge of the
instability strip. From the data~--~which was found directly from the
nonadiabatic pulsation computations~--~the blue edge could be
derived. The red edge of the instability strip is an ill-defined
quantity in astrophysics.  To have a handle what happens pulsationally
at cooler temperatures we assumed ad hoc a red line which was
displaced by a fixed $\Delta \log T_{\rm eff}$ relative to the blue
edge.  From the self-consistently determined blue edge as well as from
the intersection of the evolutionary tracks with the red line we
derived the corresponding theoretical PL relations. As a side aspect,
we looked into the variation of the pulsation constant $Q$. The
numerical results and analytical fits to them are presented in
Section~3. We compare our findings with other theoretical studies in
Section~4 and attempt to give some explanations for encountered
deviations.

\section{Computational methods and models}

Evolutionary tracks of 4, 5, 6, 7, and 10 $M_\odot$ stars were
computed with a Henyey-type stellar evolution program.  For the
stellar opacity we used the OPAL tables which were extended to the
low-temperature domain~--~below $10^4 K$~--~using the Alexander and
Ferguson (1994) data. Convective overshooting was neglected; this
aspect will be commented on in Section~4.  The computations started
always with chemically homogeneous main-sequence models. The evolution
of the model stars was followed through the helium burning so that the
major blue loops which cross the instability strip were included. To
investigate the influence of different metallicities on stellar
evolution and in particular on stellar pulsations we adopted the
following range of (X,Z) values: (0.7,0.02), (0.7, 0.004), and (0.7,
0.001).  The influence of helium abundance on stellar evolution and on
pulsational stability was studied by computing one set of stellar
masses, as listed above, with (X,Z) = (0.75, 0.004). Figure~1 shows
examples of the evolutionary tracks for stellar models with initial
homogeneous composition (X,Z) = (0.7,0.004) and for 4 and 10 $M_\odot$
with (0.75, 0.004), respectively.

Nonadiabatic radial stellar pulsation computations were {\it performed
directly on those evolutionary models\/} which are located in the
vicinity of the instability strip. This contrasts the previous
approaches where envelope models of Cepheids were computed based on
mass-luminosity relations derived from stellar evolution calculated
for other purposes. These values are kept constant throughout the
instability strip. This implies constant $M_{\rm bol}$ across the
strip.  Therefore, our approach is to be considered as more
fundamental since it is based directly on consistently computed
stellar evolution models.  Computationally, the pulsation code is of a
finite-difference relaxation type and its basics were described in
Saio et al (1983). The location of the fundamental blue edge (FBE) was
interpolated on the HR diagram between adjacent evolutionary models
for which the imaginary part of the fundamental mode changed sign from
being stable to being overstable.  The corresponding red edge (FRE)
could not be located in our computations since the indispensable
pulsation-convection interaction is not included in the code.  The
resulting lack of convective-flux leak causes the damping to be too
weak to overcome the He$^+$-partial ionization driving so that
pulsational instability extended to much too low temperatures.
Therefore, this other natural limit for pulsating stars could not be
used to study the mapping of the strip onto the P~--~L plane and to
derive lines of constant period. For these purposes, however, it is
sufficient to {\it define\/} an arbitrary line, which we chose to lie
close to the supposed FRE.  Unfortunately, the width of the
instability region is, observationally as well as theoretically, a
poorly known quantity (cf. \cite{d80}, \cite{fern90}). We adopted here
our red line to be shifted by $\Delta \log T_{\rm eff} = 0.06$ to the
red from the FBE at any luminosity. This particular choice was
motivated by the theoretical analysis of Buchler et al. (1990).  For
comparisons with observational data, we emphasize that the blue-edge
data should be considered since they depend neither on any assumptions
concerning pulsation-convection interaction nor on observational or on
any other kind of secondary calibration.

\section{Results}

In this section we present the results of radial fundamental modes at
the blue edges, at the location of our red line, and the data that can
be derived therefrom. In the figures, the blue-edge data will be
referred to by open symbols whereas red-line data will be displayed by
filled ones.  The correspondence of the different chemical
compositions with appropriate graphical symbols can be read off from
the panel inserted in Fig.~2.

Figure~1 shows~--~superposed on some evolutionary
tracks~--~the location of the blue-edge of the instability strip for
the radial fundamental mode. Despite the tracks with equal masses being
displaced considerably from each other for different chemical
compositions, the position of the FBEs is not very sensitive to
abundances and evolutionary phase.  

A linear least-square fit to the FBE data (including all different
chemical compositions) leads to the following parameterization 
$$
 \log T_{\rm eff}  = -0.036\; \log {L \over L_\odot} + 3.925\;. 
\eqno(1)
$$
For quite some purposes such a mean fit is sufficient. A closer look
at the dependencies of the FBE position on chemical composition
revealed
$$
 \Delta \log T_{\rm eff} = 0.04\;\Delta Y - 0.49\;\Delta Z\;.
\eqno(2)
$$
The coefficients were determined at $\log L / L_\odot = 3.5$.  The
sensitivity on the helium abundance has to be taken with caution as it
is based on two composition choices only: (X,Z)=(0.7,0.004) and
(0.75,0.004), respectively.  Our helium dependence is indeed
considerably weaker than the ones given by Iben and Tuggle (1972) or
Chiosi et al. (1993).  Nevertheless, the compensating action of
$\Delta Y$ and $\Delta Z$ remains unchanged.

When the periods of the fundamental mode of the stellar models at the
FBE are plotted versus their luminosity we arrive directly at the PL
relation. The same procedure was applied along the red line of the
instability strip. The results for the blue edge (open symbols) and
the red line (filled symbols) are shown in Fig.~2.  Only crossings of
the instability strip during the Cepheid blue-loops (cf. \cite{kw90})
were considered.  We notice no systematic difference neither in slope
nor in zero-point of the PL relation on metallicity or helium
abundance. With good accuracy, the red-line data are shifted by a
constant amount relative to the blue-edge data.  Again, no systematic
dependence of the corresponding PL relation on metallicity or helium
was found.

To quantify the PL relation, we performed a least-square straight-line
fit through the blue-edge and the red-line data separately. Again,
only second and later crossings of the instability strip are included
in the data.  For the blue edge we found
$$
	\log {L \over L_\odot} = 2.573 + 1.270\; \log P
\eqno(3)
$$
with a mean $\log L / L_\odot$ deviation from the fit of $3.6
\cdot 10^{-2}$.  The red-line data can be described by
$$
	\log {L \over L_\odot} = 2.326 + 1.244\; \log P\;, 
\eqno(4)
$$ 
for which the mean $\log L / L_\odot$ deviation from the fit
amounts to $3.75 \cdot 10^{-2}$.  The slopes of both edges deviate
from each other by about 10 \%.  This means that the constant width in
$\log T_{\rm eff}$ of the instability strip translates into a relation
of only slightly variable width on the $\log P$~--~$\log L$ plane. The
zero-points of the PL relation shift by $\Delta \log L/L_\odot = 0.273$
at $\log P = 1.0$.

It is worth noticing that the PL relation hardly changes even if the
first crossing of the instability strip (before the onset of core
helium burning) of the stellar models is included in the fits to the
PL relation.

When computing the nonadiabatic fundamental modes for our stellar
models we also obtained the corresponding pulsation constants $Q
\equiv P \cdot \sqrt(\rho/\rho_\odot)$. This quantity was frequently
used in the past in qualitative semi-analytical arguments on the PL
relation of Cepheids in the literature, therein it was usually treated
as a constant. Since stellar evolution during the Cepheid phase cannot
be mimicked by simple homology transformations between different
evolutionary phases within the instability strip we quantify its
variation along the instability strip and for the masses and
luminosities of Cepheids.

Figure~3a displays the nonadiabatically derived pulsation constants as
a function of effective temperature of the blue-edge and the red-line,
respectively. Evidently, $Q$ is not constant. Connecting open and
filled symbols belonging to the same branch of an evolutionary track,
we find $Q$ to increase slightly when a star of a given mass crosses
the instability strip from blue to red. The largest variation of $Q$,
by roughly 50 \%, occurs upon increasing luminosity along the
instability strip. As already observed in the PL relation, also the
$Q$-values do not depend on particular choices of metallicity or
helium abundance.  When plotting the $Q$-values versus period the
blue-edge~--~red-line offset which is seen in Fig.~3a reduces
considerably (cf. Fig.~3b). Only a slight vertical shift of at most
0.005 in $Q$ remains at a given period between blue-edge and red-line
data. The following quadratic fits the data satisfactory:
$$
 Q({\rm d}) =   3.47\cdot 10^{-2} 
              + 5.2 \cdot 10^{-3}  \log P({\rm d}) 
              + 2.8 \cdot 10^{-3} [\log P({\rm d})]^2\;.
\eqno (5)
$$

Our theoretical data delineating the positions of the FBE, the red
line of the strip, and the associated PL relation allow the
computation of lines of constant period along the instability strip on
the HR diagram.  Since we restricted ourselves to the computation of
periods at the FBE and along the red line, our loci of constant period are
straight. This must, however, not necessarily be so. We found that the
slope is a function of the position~--~parameterized by the
period~--~in the instability strip. It can be approximated by the
following linear relation:
$$
      \left ( { \partial \log L / L_{\odot} 
               \over
                \partial \log T_{\rm eff}
              }
      \right )_P
	= 4.80 + 0.608 \log P\;.
\eqno (6)
$$
The slope changes therefore by 20 \% when going from  $P = 1$~day to 
$P = 40$~days.
The existence of the slope change is equivalent to the fact that 
the strength of the temperature dependence of Q-value depends
on period as seen in Fig. 3b.
If we integrate equation (6), we see that the coefficient of the 
color term in a period-luminosity-color (PLC) relation should be a 
function of period.
When PLC relations are derived from observational data, however,
it is always assumed that the coefficient of the color
correction term is independent of period.
This assumption is, strictly speaking, inconsistent 
with our theoretical results.
We are not aware of any
attempt to take this effect into consideration and we hypothesize that
this might be the reason for the failure of establishing a PLC
relation for Galactic Cepheids (cf. Fernie 1990).

\section{Discussion}

According to our evolution/stability computations, incorporating new
OPAL opacity data and low-temperature extensions provided by Alexander
and Ferguson (1994) we find {\it no\/} significant metallicity and no
helium dependence of the $PL$ relation of Cepheids. Our results apply
to bolometric quantities, however. Any observational realization of
the PL relation such as e.g. a $\log P-M_{\rm V}$ relation will depend
to some degree on the the particular chemical composition of the
Cepheids through the metallicity sensitivity of the bolometric
correction.  In fact, the metallicity dependence of Stothers'(1988)
$P - M_V$ relation (which is weak though) mostly comes from the 
bolometric correction.
The magnitude of the metallicity dependence of the bolometric correction
is dictated by the
choice of the filter passband.  The quantification of this effect is
not the purpose of this paper and has to be addressed by the theory of
stellar atmospheres.

For the blue-edge and the red-line data, the slopes of the PL relation are
1.270 and 1.244, respectively. Our FBE slope is similar to what Chiosi
et al.~(1993) (1.280) found but larger than the 1.19 which was given
as a mean value by Stothers (1988). From observational data, Cox
(1980) derived a slope of the bolometric PL relation of 1.15. This PL
relation was based on bolometrically corrected $M_{\rm V}$
data. Stothers (1988) refers to the infrared PL relation which should
be close to the bolometric one and cites slopes between 1.16 and 1.39.
We conclude from this that the slopes of our PL relations lie within
the range of previous theoretical ones. Therefore, the latest
improvements in opacities do not influence the {\it slope\/} of the
Cepheid PL relation.  

Compared with Chiosi et al.'s (1993) mild overshooting case our
zero-point (measured at $P=10$ days) is brighter by 0.02 in $\log
L/L_\odot$. The particular treatment of overshooting influences mainly
the ML relation. Compared with the ML relation the luminosity shift
due to overshooting is about a factor of five less in the PL
relation. The slope of the PL relation is even more stable with regard
to changes in the treatment of overshooting. 
In view of the experience of Chiosi et al. (1993) we did not further
elaborate here on computing series of stellar models with different
overshooting parameterizations.

The present physical calibration is slightly brighter than
observational ones. Using Galactic Cepheids as calibrators gives
for a ridge-line Cepheid with $P = 10$~d $M_{\rm V} = -4.20$
(Sandage and  Tammann 1968). Based on an adapted LMC modulus of 
($m-M$) = 18.57, Madore and Freedman (1991, 1997) find a corresponding
value of $M_{\rm V} = -4.21$. Using stellar radii as calibrators
Di~Benedetto (1997) finds $M_{\rm V} = -4.10$. A 10-days Cepheid
with $\log T_{\rm eff} = 3.76$ has a bolometric correction of 
BC$\approx 0.13$ (Lang 1991). With this, one obtains $M_{\rm
bol}(P=10~{\rm d}) \approx -4.30$. Our physical calibration gives
for a corresponding Cepheid on the ridge line (half-way between the FBE
and our red line)  $M_{\rm bol}(P=10~{\rm d}) = -4.50$. 

The steeper slopes of the theoretical PL relations and the shifts in
the zero points, say at 10 days period, could be understood by noticing
that the Cepheid population of the instability strip is not
uniform. Fernie (1990) showed evidence that Cepheids favor to populate
regions close to the blue-edge at lower luminosities and close to the
red line a higher luminosities.  A ridge line determined from an
arbitrary sample of Cepheids neglecting their inhomogeneous
distribution across the instability strip will lead to a PL relation
with a shallower slope than what results from an instability strip
with a homogeneous distribution of variable stars.  If this
inhomogeneity is not accounted for properly in observational data a
shallower `ridge-line' PL relation results when compared with the
theoretically predicted one.

Finally we quantified the variation of the pulsation constant $Q$
across and along the instability strip as far as it is relevant for
Cepheid studies. The variation is considerable as a function of
luminosity: Going from 4 to 10 $M_\odot$, $Q$ changed by roughly 50
\%.  An appropriate fit was presented in Sect.~3.  An important result
is that the $Q$ values change very little as helium or heavy-element
abundances are changed. Furthermore, we showed how the lines of
constant period change their slopes (when they are assumed to be
straight lines across the instability strip) as a function of
period. The fit to the data allowed a quantification of the effect so
that it could be incorporated into semi-analytical discussions of
pulsation properties of Cepheids.

Acknowledgments: We are grateful to G.A. Tammann and A.R. Sandage for
encouraging this study.  H.S. acknowledges financial support from the
Japanese Society for Promotion of Science and the Swiss National
Science Foundation.  A.G. was financially supported by the Swiss
National Science Foundation through a PROFIL2 fellowship.

\clearpage

\clearpage

\begin{figure}
\plotone{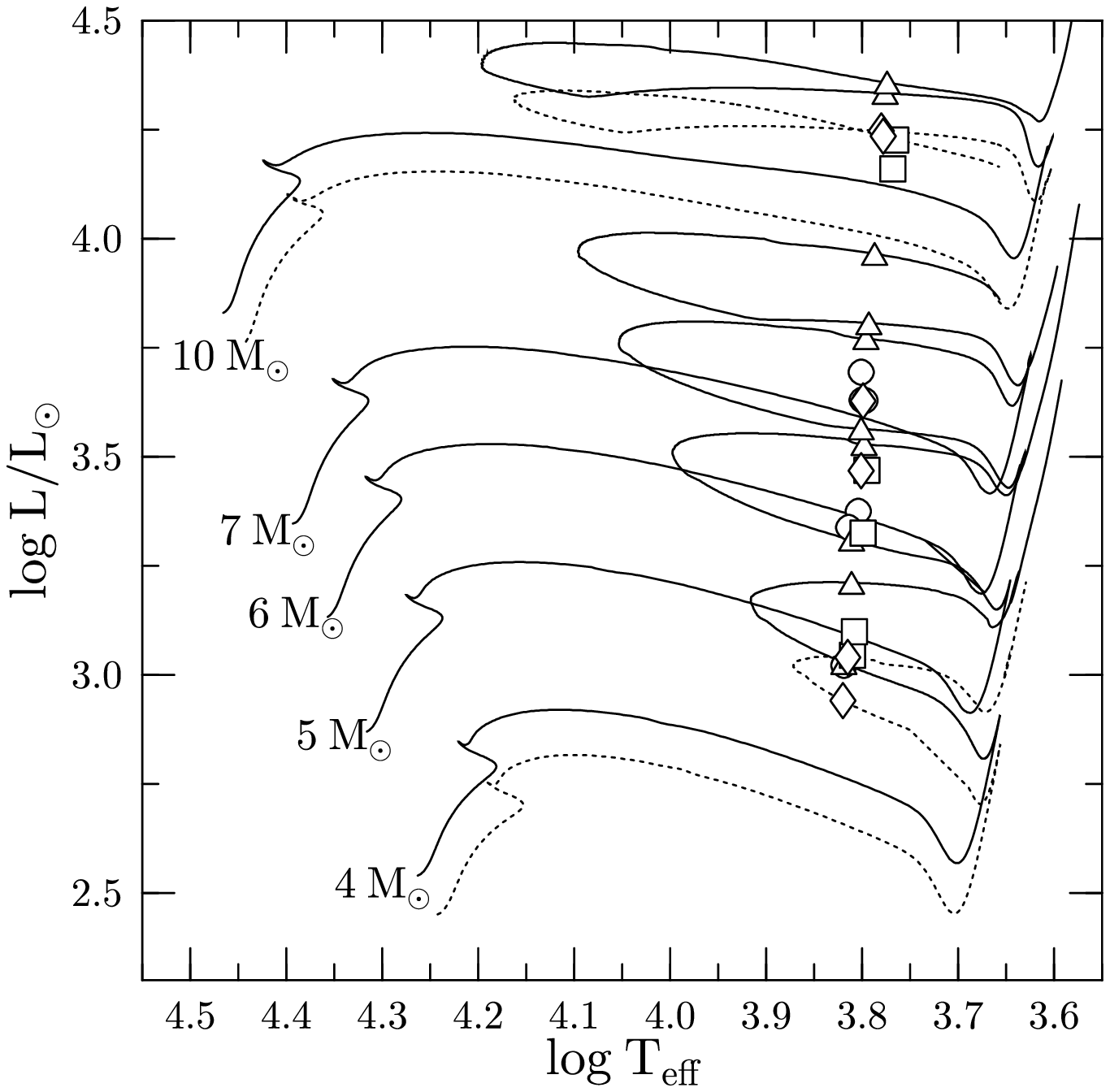}
\caption{Evolutionary tracks of various stellar models on
           the HR diagram. The continuous line shows the (X,Z) = (0.7, 0.004)
           composition;  dotted lines trace out tracks of two stellar 
           masses with (X,Z)=(0.75,0.004).  
           Open symbols indicate the position of
           the FBE.  The different symbol-types stand for the
           different chemical compositions (see the panel inserted in
           Fig.~2) \label{fig1}}
\end{figure}

\clearpage

\begin{figure}
\epsscale{.75}
\plotone{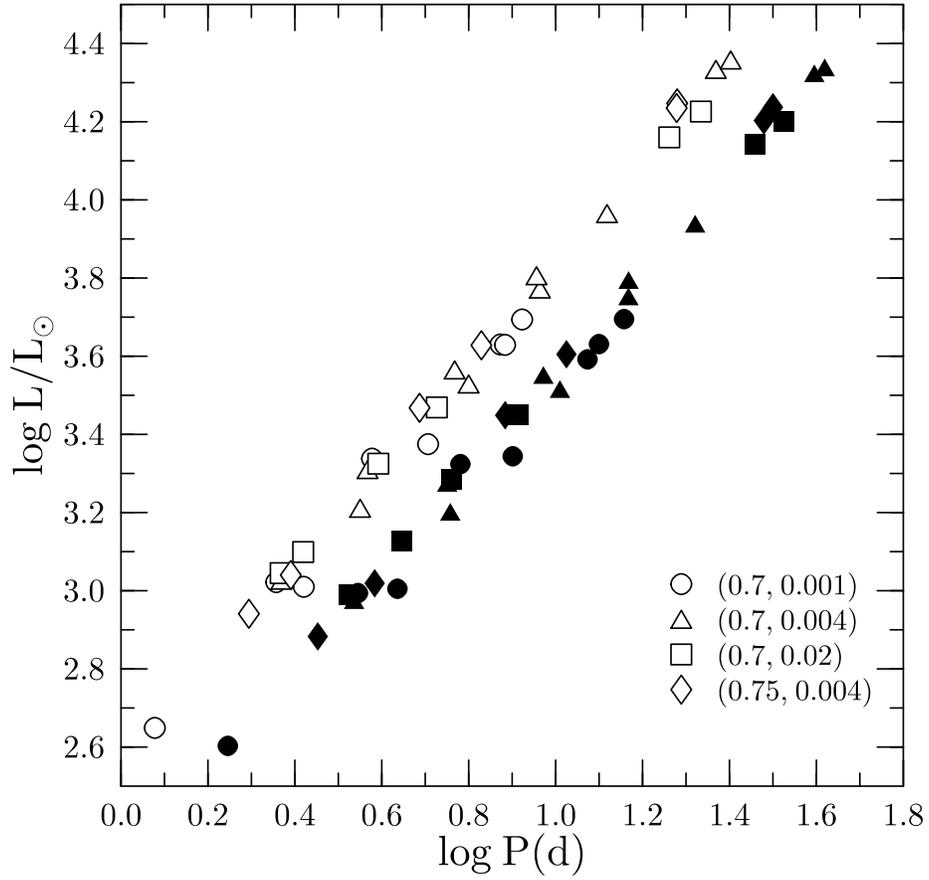}
\caption{Period-luminosity relation for the blue edge (open
         symbols) and for the ad hoc defined red line (filled symbols). All
         different chemical compositions are shown in the same graph.
	 \label{fig2}}
\end{figure}

\clearpage

\begin{figure}
\epsscale{.5}
\plotone{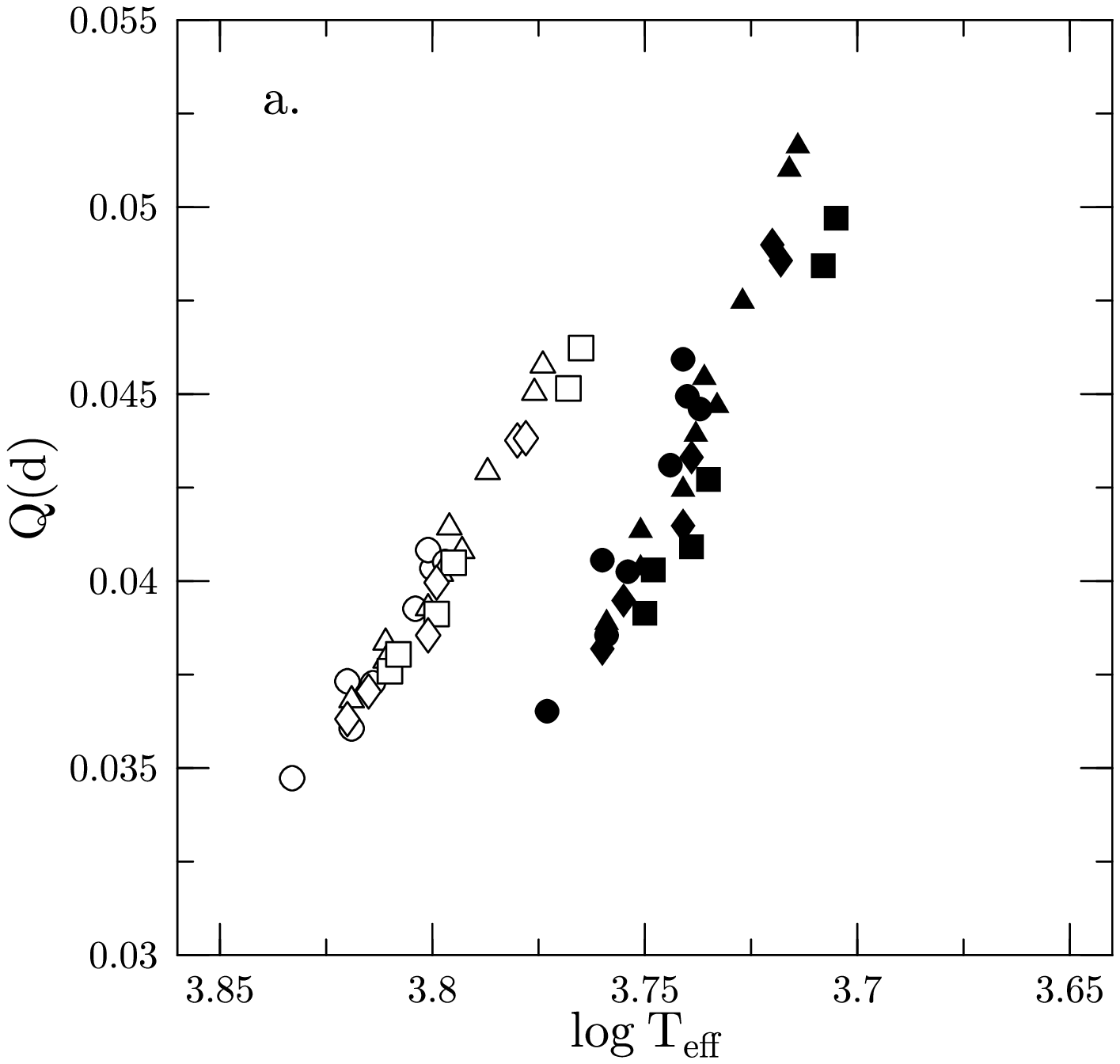}
\plotone{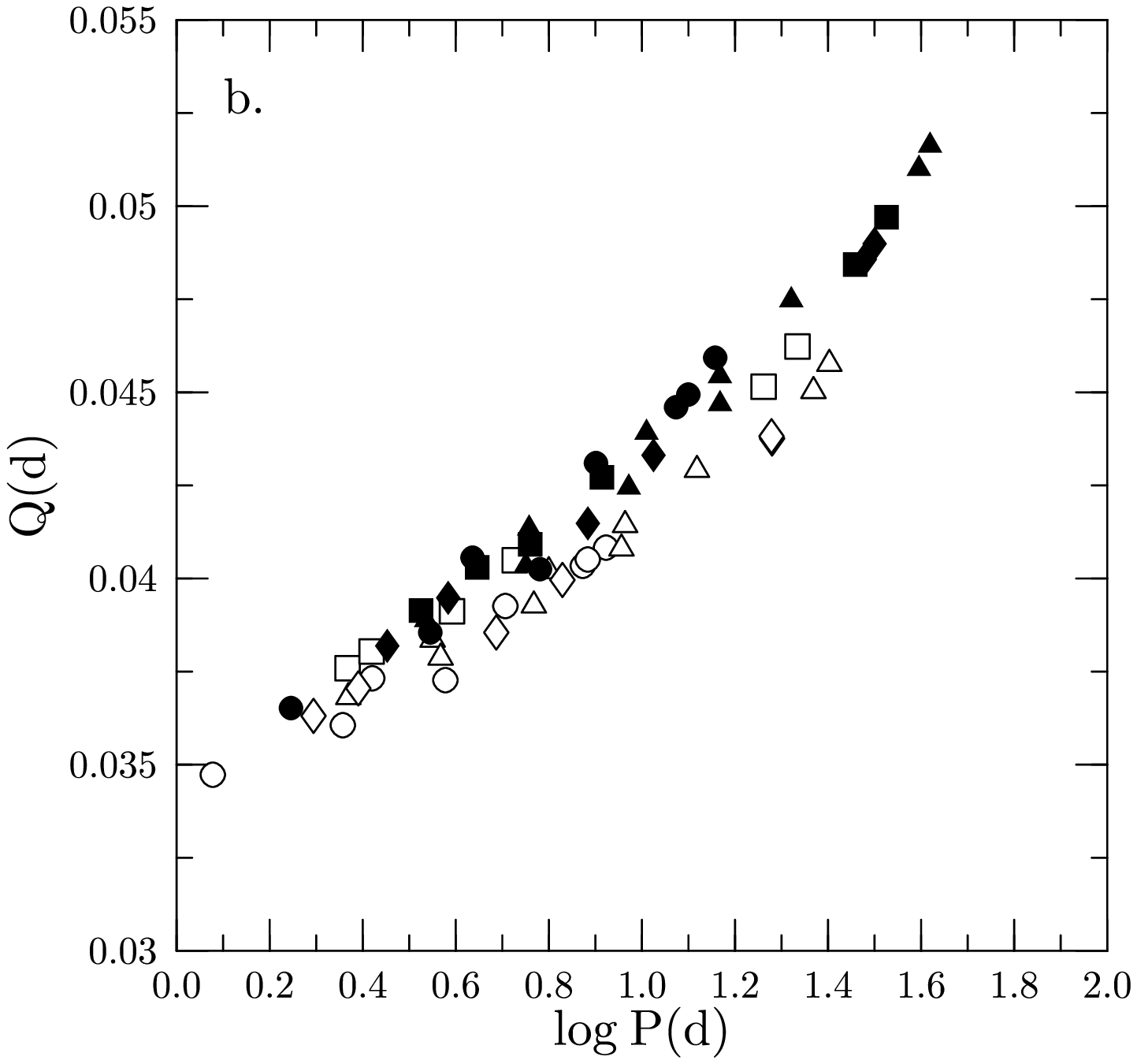}
\caption{{\bf a:} Pulsation constant $Q$ as a function 
           of effective temperature. Open symbols indicate blue-edge 
           values whereas filled symbols stand for the corresponding 
           red-line data.
           {\bf b:} Pulsation constant $Q$ as a function of the
           logarithm of the Period.
	   Blue-edge and red-line data converge nicely so that the
           period dependence can be fitted with a simple quadratic
	   function. Details are given in the text. \label{fig3}}
\end{figure}


\end{document}